\documentclass[reprint,amsmath,amssymb,aps,pra]{revtex4-1}
\usepackage{lmodern}

\usepackage[T1]{fontenc}
\usepackage[latin9]{inputenc}
\usepackage{amsmath}
\usepackage{graphicx}

\makeatletter
\usepackage{dcolumn}
\usepackage{bm}

\usepackage{physics}
\usepackage{wasysym}

\makeatother

\begin{document}
\title{Scalable collective Lamb shift of a 1D superconducting qubit array
in front of a mirror}
\author{Kuan-Ting Lin$^{1}$, Ting Hsu$^{1}$, Chen-Yu Lee$^{1}$ , Io-Chun
Hoi$^{2}$, and Guin-Dar Lin$^{1}$}
\affiliation{$^{1}$Center for Quantum Science and Engineering, Department of Physics,
National Taiwan University, Taipei 10617, Taiwan}
\affiliation{$^{2}$Center for Quantum Technology and Department of Physics, National
Tsing Hua University, Hsinchu 30013, Taiwan }
\begin{abstract}
We theoretically investigate resonant dipole-dipole interaction (RDDI)
between artificial atoms in a 1D geometry, implemented by $N$ transmon
qubits coupled through a transmission line. Similarly to the atomic
cases, RDDI comes from exchange of virtual photons of the unexcited
modes, and causes the so-called collective Lamb shift (CLS). To probe
the shift, we effectively set one end of the transmission line as
a mirror, and examine the reflection spectrum of the probe field from
the other end. Our calculation shows that when a qubit is placed at
the node of the standing wave formed by the incident and reflected
waves, even though it is considered to be decoupled from the field,
it results in large energy splitting in the spectral profile of a
resonant qubit located elsewhere. This directly signals the interplay
of virtual photon processes and explicitly demonstrates the CLS. We
further derive a master equation to describe the system, which can
take into account mismatch of participating qubits and dephasing effects.
Our calculation also demonstrates the superradiant and subradiant
nature of the atomic states, and how the CLS scales when more qubits
are involved.
\end{abstract}
\maketitle

\section{Introduction}

One of the intriguing phenomena of quantum electrodynamics is the
emergence of the Lamb shift, which was first discovered by Lamb in
1947 \citep{Lamb1947}, corresponding to the energy difference between
2S$_{1/2}$ and 2P$_{1/2}$ levels of a hydrogen atom. The understanding
of such a shift opened up a new chapter of physics now known as quantum
field theory, bringing in a concept that quantum vacuum must be treated
as a zero-point state of numerous harmonic oscillators (photon modes),
and quantum fluctuations allow both real and virtual processes to
have physical effects. This perspective of quantum vacuum also plays
an essential role in various scenarios such as spontaneous decay emission,
squeezed vacuum states \citep{Carmichael1987,Schiller1996}, and the
Casimir effect \citep{Casimir1948,Bordag2001,Wilson2011}. Recently,
resonant dipole-dipole interaction (RDDI) mediated via exchange of
virtual photons between multiple atoms has become one of the most
interesting topics in different contexts \citep{LeKien2005,Zhu2016,Bromley2016,Shahmoon2017,Perczel2017,Beterov2018}.
Such RDDI results in the collective version of Lamb shift, sometimes
also termed the cooperative Lamb shift (CLS) due to its close connection
to cooperative phenomena such as super- and subradiance \citep{Arecchi1970,Friedberg1973,Scully2009}.
For past few years, CLS regarding atomic systems have been experimentally
demonstrated and studied in atomic clouds \citep{Araujo2016,Roof2016},
in nano-layer gases \citep{Keaveney2012,Peyrot2018}, ensembles of
embedded nuclei \citep{Roehlsberger2010}, and trapped ions \citep{Meir2014}.
Main challenges of observing CLS in atomic systems originate from
vacuum mediated coupling weakened very fast as separation increases
in 3D space. In order to probe the shift, ideally atoms must be placed
at a distance comparable to the transition wavelength, or inside cavities
or waveguides where field can be confined or directed , thus enhancing
the interaction strength. Based on such consideration, it is suggested
that the circuit quantum electrodynamical (circuit QED, or cQED) systems
are a perfect testbed for observing cooperative phenomena.

\begin{figure}
\begin{centering}
\includegraphics[width=8cm]{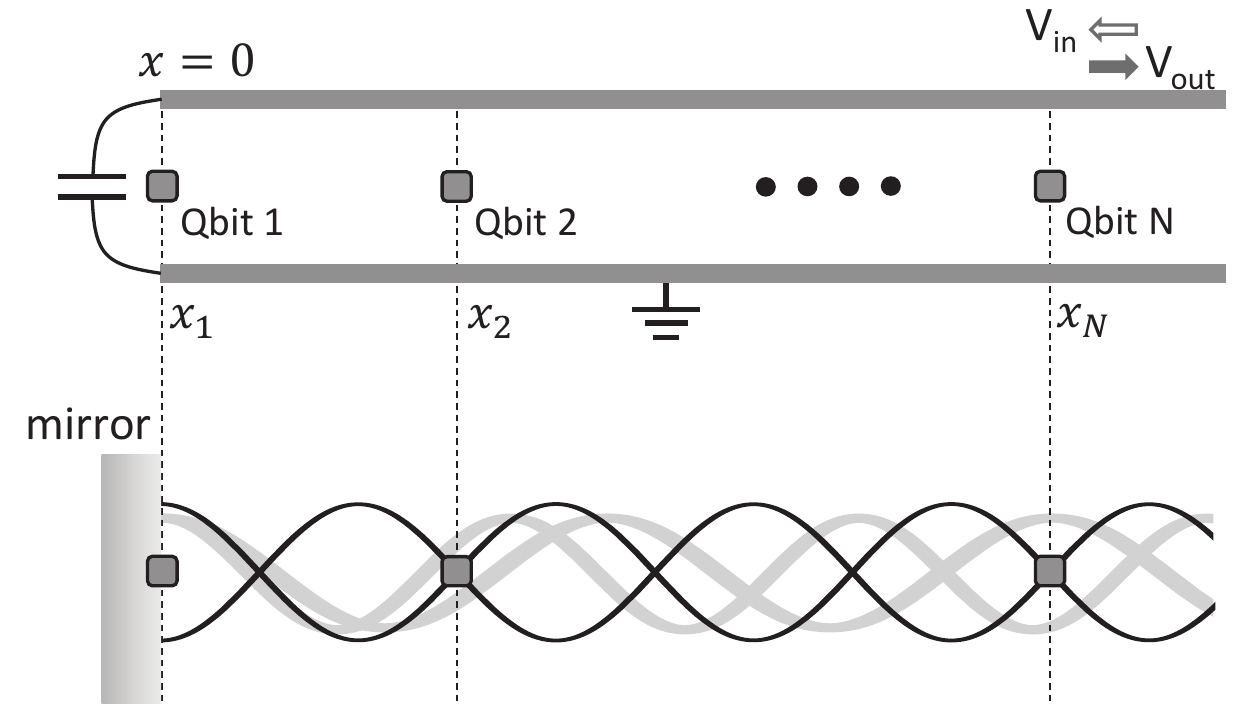}
\par\end{centering}
\caption{Architecture of the 1D array of transmon qubits coupled through a
microwave waveguide, whose one end is terminated by a large capacitor
at $x=0$, effectively serving as an antinode mirror. The probe field
is fed from the other end of the waveguide, coherently superposes
with the reflected field, forming a standing wave. When other qubits
are placed at the nodes, they do not directly interact with probe
photons. However, the qubits can still couple to other vacuum modes
of continuous spectrum, mediating the RDDI only through virtual processes.
\label{fig:architecture}}
\end{figure}
Circuit QED systems deal with superconducting artificial atoms coupled
through 1D microwave waveguides. They are relatively more easily fabricated
to achieve the strong coupling or the superradiant regime compared
to the atomic counterpart \citep{Gu2017}, and have been used extensively
to study the Tavis-Cummings model \citep{Fink2009}, dipole-dipole
coupling \citep{Filipp2011}, photon-ensemble interaction, super-
and subradiance \citep{Lalumiere2013,Loo2013,Mlynek2014,Mirhosseini2018,Lambert2016},
and quantum information oriented applications \citep{Facchi2016,Liao2018}.
Up to present, the observation of CLS in cQED systems is still scarce
except for \citep{Loo2013}, where two qubits are both pumped in an
open waveguide, resulting in collective decay linewidth larger than
the shift, seriously degrading the visibility of CLS. We here study
the emergence of CLS when a series of transmon qubits are arranged
in front of a mirror, and probed for the reflection spectrum. In order
to identify RDDI that is exclusively from mediation of virtual photons,
we put one qubit at an antinode while others at nodes with respect
to their transition wavelength as shown in Fig. \ref{fig:architecture}.
Intuitively speaking, since these qubits at nodes do not directly
interact with the probing field, they have no effect on the antinode
qubit's spectral profile through real photon exchange. This is however
not the entire story because one neglects contributions from the whole
range of vacuum modes, which, except for the resonant one, are not
physically excited but in fact responsible for RDDI. The advantage
of insertion of a mirror is to introduce destructive interference
that suppresses the collective decay linewidth, hence improving the
visibility of the CLS.

This paper is organized as follows: Sec. \ref{sec:Model} presents
a cQED system of a half-infinite waveguide, and derives the model
for RDDI associated with it. Sec. \ref{sec:2atomcase} mainly discusses
two-qubit cases, focusing on the superradianct condition and emergence
of CLS. We calculate the reflection spectrum, commonly measured in
current experiments \citep{Hoi2015,Wen2018,Wen2019}, showing the
observable CLS splitting in the profile. We also study the dephasing
and power broadening effects as well as level anharmonicity for real
transmon artificial atoms. Sec. \ref{sec:manyatom} discusses multi-qubit
cases, demonstrating the scaling law for the CLS and number of qubits.
To explain this we present an effective reduced scheme. Finally we
conclude this work in Sec. \ref{sec:conclusion}.

\section{Model\label{sec:Model}}

\subsection{Dipole-dipole interaction and master equation\label{subsec:rddi_mastereq}}

We consider a linear chain of $N$ transmon qubits coupled to a common
1D waveguide whose one end is terminated by a very large capacitor.
This amounts to set the end as an antinode mirror regarding the standing
waves in such an architecture. Different from a discrete spectrum
in a cavity case with two mirrors, our system allows continuum of
the photon modes. The Hamiltonian describing this system can be written
as $H=H_{S}+H_{B}+H_{int}$ \citep{Koch2007,Peropadre2013,Dorner2002,Hoi2015}
with the atomic part $H_{S}=\sum_{i}\hbar\omega_{i}\sigma_{i}^{+}\sigma_{i}^{-}$,
the field part $H_{B}=\int_{0}^{\infty}\hbar\omega a_{\omega}^{+}a_{\omega}d\omega,$
and the interaction under the rotating wave approximation
\begin{equation}
H_{int}=i\sum_{i}\int_{0}^{\infty}d\omega\hbar g_{i}(\omega)\cos(k_{\omega}x_{i})a_{\omega}\sigma_{i}^{+}+H.c..\label{eq:interaction}
\end{equation}
Here, $\omega_{i}$ denotes the transition frequency between the excited
state $|e\rangle_{i}$ and the ground one $|g\rangle_{i}$ of the
$i$th qubit located at $x_{i}$, and $\sigma_{i}^{+}=|e\rangle_{i}\langle g|$
and $\sigma_{i}^{-}=|g\rangle_{i}\langle e|$ represent its raising
and lowering operators, respectively. The operator $a_{\omega}^{\dagger}$
($a_{\omega}$) creates (annihilates) a photon of frequency $\omega$,
whose mode function is of the form $\sim\cos k_{\omega}x$ due to
presence of the antinode mirror at $x=0$. The wavenumber $k_{\omega}=\omega/v$
with $v$ the speed of light in the waveguide. Note that $a_{\omega}^{+}$
and $a_{\omega}$ satisfy the commutation relation $[a_{\omega^{'}},a_{\omega}^{+}]=\delta(\omega-\omega^{'})$.
Following the standard procedure to trace out the photonic degrees
of freedom \citep{Lehmberg1970} and applying the Born-Markov approximation,
we arrive at the master equation
\begin{equation}
\begin{aligned}\frac{d\rho}{dt} & =i\sum_{i}\delta_{i}[\sigma_{i}^{+}\sigma_{i}^{-},\rho]\\
 & -i\sum_{ij}(\Delta_{ij}^{+}-i\gamma_{ij}^{-})[\sigma_{i}^{+}\sigma_{j}^{-},\rho]\\
 & +i\sum_{i}\Omega_{p}^{i}\cos(k_{p}x_{i})[\sigma_{i}^{+}+\sigma_{i}^{-},\rho]\\
 & +\sum_{ij}(\gamma_{ij}^{+}+i\Delta_{ij}^{-})\mathcal{L}_{ij}[\rho]\\
 & +\sum_{i}\gamma_{i}^{\phi}\mathcal{L}_{i}^{\phi}[\rho].
\end{aligned}
\label{eq:master eq}
\end{equation}
In this master equation, we have explicitly included a continuous-wave
probe field incident from the other end of the waveguide with a detuning
$\delta_{i}=\omega_{p}-\omega_{i}$, the associated Rabi frequency
$\Omega_{p}^{i}$ seen by the $i$th qubit, and a wavenumber $k_{p}=\omega_{p}/v$.
The superoperator $\mathcal{L}_{ij}[\rho]\equiv2\sigma_{j}^{-}\rho\sigma_{i}^{+}-\sigma_{i}^{+}\sigma_{j}^{-}\rho-\rho\sigma_{i}^{+}\sigma_{j}^{-}$
deals with self and cooperative dissipative processes. And $\mathcal{L}_{i}^{\phi}[\rho]\equiv2\sigma_{i}^{ee}\rho\sigma_{i}^{ee}-\sigma_{i}^{ee}\rho-\rho\sigma_{i}^{ee}$
with $\sigma_{i}^{ee}=|e\rangle_{i}\langle e|$ is added by hand to
account for individual pure dephasing characterized by $\gamma_{i}^{\phi}$.
The dipole-dipole interaction, obtained by summing all contributions
from the photon mode continuum, is now contained in $\gamma_{ij}^{\pm}=(\gamma_{ij}\pm\gamma_{ji})/2$
and $\Delta_{ij}^{\pm}=(\Delta_{ij}\pm\Delta_{ji})/2$ with
\begin{equation}
\gamma_{ij}=\frac{\gamma_{ij}^{0}}{2}\left[\cos k_{j}(x_{i}+x_{j})+\cos k_{j}\left|x_{i}-x_{j}\right|\right]\label{eq:gnm}
\end{equation}
\begin{equation}
\Delta_{ij}=\frac{\gamma_{ij}^{0}}{2}\left[\sin k_{j}(x_{i}+x_{j})+\sin k_{j}\left|x_{i}-x_{j}\right|\right],\label{eq:dnm}
\end{equation}
where $\gamma_{ij}^{0}\equiv\sqrt{\gamma_{i}(\omega_{j})\gamma_{j}(\omega_{j})}$
with the bare decay rate $\gamma_{i}=\pi g_{i}^{2}(\omega_{j})$ evaluated
at the $j$th qubit's transition frequency $\omega_{j}$ (see Appendix
A for details).

Here are a few remarks regarding the forms of Eqs. (\ref{eq:gnm})
and (\ref{eq:dnm}). First, for an open waveguide without a mirror,
it can be proven that the dipole-dipole interaction between the $i$th
and $j$th qubits depends only on the relative distance $\left|x_{i}-x_{j}\right|$
\citep{Gu2017}. The mirror effectively places image atoms on the
other side of the mirror. Therefore qubit $i$ does not only see the
real qubit $j$ at a distance $\left|x_{i}-x_{j}\right|$ but also
the image one at distance $(x_{i}+x_{j})$. Secondly, these rates
and shifts are related through the Kramers-Kronig (KK) relations
\begin{align}
\gamma_{ij}= & \frac{1}{\pi}\int\frac{\Delta_{ij}(\omega^{\prime})}{\omega-\omega^{\prime}}d\omega^{\prime}\\
\Delta_{ij}= & -\frac{1}{\pi}\int\frac{\gamma_{ij}(\omega^{\prime})}{\omega-\omega^{\prime}}d\omega^{\prime}.
\end{align}
Note that, in general, $\gamma_{ij}^{\pm}$ and $\Delta_{ij}^{\pm}$
are non-zero with non-identical qubits, leading to non-Lindblad behavior
\citep{Dung2002}. For identical qubits where the sub-indices are
interchangeable, $\gamma_{ij}^{-}$ and $\Delta_{ij}^{-}$ vanish
and hence the master equation retains the Lindblad form; $\Delta_{ij}$
then directly contributes to the CLS.

\subsection{Scattering and Reflection\label{subsec:reflection}}

In order to probe the CLS configuration, we manage to feed the probe
signal from and acquire its reflection spectrum on the open end. As
measured in many experiments \citep{Koshino2012,Peropadre2013,Hoi2015,Wen2018},
the reflection coefficient is obtained by
\begin{equation}
r(x,t)\equiv\left|\left\langle V_{out}(x,t)/V_{in}(x,t)\right\rangle \right|,\label{eq:reflection}
\end{equation}
where the output signal $V_{out}(x,t)=V_{in}(x,t)+V_{sc}(x,t)$ with
the input voltage $V_{in}$ and scattered one $V_{sc}$. The input
signal is assumed to be of the form 
\begin{equation}
V_{in}(x,t)=V_{0}e^{ik_{p}r}\label{eq:vin}
\end{equation}
viewed from the rotating frame of the probe frequency, where $V_{0}$
is the amplitude of the input voltage with its corresponding wave
number $k_{p}$. The scattered voltage can be calculated from the
flux \citep{Peropadre2013,Yurke1984}
\begin{equation}
\begin{aligned}\Phi(x,t) & =\sqrt{\frac{\hbar Z_{0}}{\pi}}\int\frac{d\omega}{\sqrt{\omega}}\cos k_{\omega}x(a_{\omega}+a_{\omega}^{\dagger})\end{aligned}
\label{eq:flux}
\end{equation}
with the characteristic impedance $Z_{0}$. Then the scattered signal
is obtained by differentiating the positive frequency part $V_{sc}=\partial\Phi^{out}/\partial t$
for the outgoing wave. In the probe-frequency frame,
\begin{equation}
V_{sc}(x,t)=-i\sqrt{\frac{\hbar Z_{0}}{4\pi}}\int_{0}^{\infty}\sqrt{\omega}\tilde{a}_{\omega}(t)e^{ik_{\omega}x-i(\omega-\omega_{p})t}d\omega.\label{eq:vsc}
\end{equation}
Here we have used the fact that the field operator can be expressed
in terms of the slowly-varying amplitude $a_{\omega}(t)=\tilde{a}_{\omega}(t)e^{-i\omega t}$
and $\dot{\tilde{a}}_{\omega}\approx0$. Through the standard procedures,
as summarized in Appendix A, the photonic operator is related to the
atomic one \citep{Carmichael2003,Scully1997}
\begin{equation}
\tilde{a}_{\omega}(t)=-\sum_{i=1}^{N}g_{i}(\omega)\int_{0}^{t}\widetilde{\sigma}_{i}^{-}(t^{\prime})e^{i(\omega-\omega_{i})t^{\prime}}dt^{\prime},\label{eq:field op}
\end{equation}
where the atomic operator is also assumed of the form $\sigma_{i}^{-}(t)=\tilde{\sigma}_{i}^{-}(t)e^{-i\omega_{i}t}$.
Substituting Eq. (\ref{eq:field op}) into Eq. (\ref{eq:vsc}), and
using Eqs. (\ref{eq:reflection}) and (\ref{eq:vin}), we then have
the scattered signal and the reflection coefficient, respectively,
\begin{equation}
V_{sc}=i\sum_{i=1}^{N}\sqrt{\hbar\pi Z_{0}\omega_{i}}g_{i}(\omega_{i})\tilde{\sigma}_{i}^{-},\label{eq:form vsc}
\end{equation}
\begin{equation}
r=\left|1+i\sum_{i=1}^{N}\sqrt{\hbar\pi Z_{0}\omega_{i}}g_{i}(\omega_{i})\cos k_{p}x_{i}\left\langle \sigma_{i}^{-}\right\rangle /V_{0}\right|.
\end{equation}
The photon-atom coupling strength for transmon qubits is given by
\begin{align}
g_{i}(\omega) & =e\beta_{i}\left(\frac{E_{J}^{(i)}}{8E_{C}^{(i)}}\right)^{1/4}\sqrt{\frac{2Z_{0}\omega}{\pi\hbar}},
\end{align}
where $e$ is the electron charge; $Z_{0}$ is the characteristic
impedance of the transmission line; $\beta_{i}=C_{C}^{i}/C_{T}^{i}$
is the ratio between the capacitor $C_{C}^{i}$ of the transmission
line and the total capacitor $C_{T}^{i}$; $E_{J}^{(i)}$ and $E_{C}^{(i)}$
are the Josephson energy and the charging energy, respectively, of
the $i$th qubit \citep{You2011,Koch2007,Devoret2004}. Note that
the input voltage $V_{0}$ is viewed right outside the outmost qubit
(the $N$th one), and is connected to the Rabi frequency via
\begin{equation}
V_{0}=\frac{\Omega_{p}^{N}}{2g_{N}(\omega)}\sqrt{\frac{\hbar Z_{0}\omega}{\pi}}.\label{eq:v0 omegap}
\end{equation}
By expressing $V_{0}$ in terms of $\Omega_{p}$, we finally obtain
the reflection coefficient

\begin{equation}
r=\left|1+i\sum_{i=1}^{N}\frac{2\eta_{Ni}\gamma_{i}}{\Omega_{p}^{N}}\cos k_{p}x_{i}\left\langle \sigma_{i}^{-}\right\rangle \right|,\label{eq:final form reflection}
\end{equation}
with $\eta_{Ni}=(E_{J}^{(N)}E_{c}^{(i)}/E_{J}^{(i)}E_{c}^{(N)})^{1/4}\beta_{N}/\beta_{i}$.
The atomic variables $\langle\sigma_{i}^{-}\rangle$ needs to be solved
by evaluating the master equation (\ref{eq:master eq}), which can
be done numerically, or analytically only under the weak field approximation
for the steady state. We will discuss the results in the following
sections.

\section{Superradiance and collective Lamb shift for two-atom cases\label{sec:2atomcase}}

\subsection{Reflection spectrum }

\begin{figure}
\begin{centering}
\includegraphics[width=8.5cm]{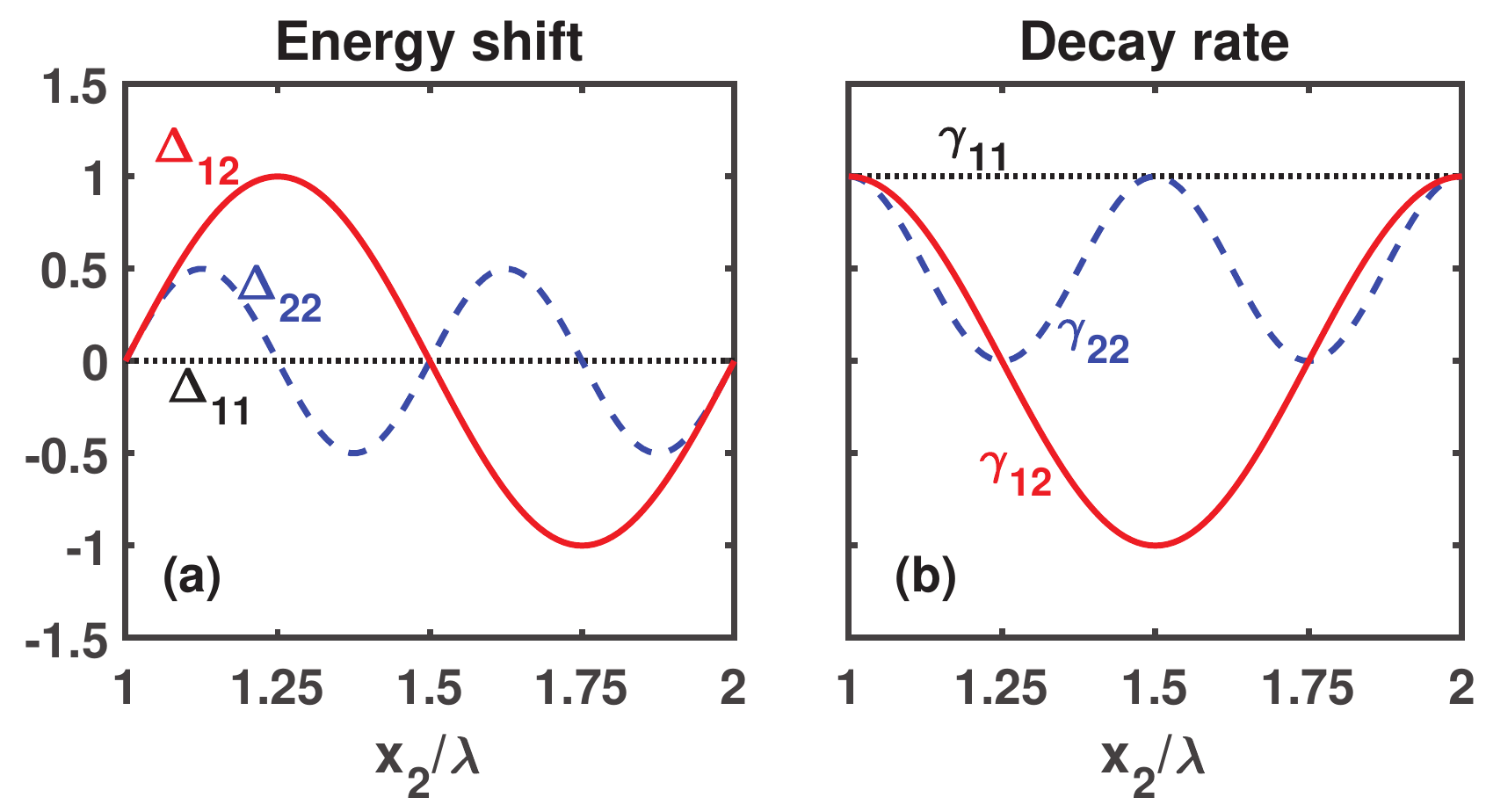}
\par\end{centering}
\caption{Energy shifts and decay rates (in units of $\gamma_{0}$) of the two-qubit
system for varied $x_{2}$. $\Delta_{ii}$ and $\gamma_{ii}$ represent
the single-atom energy shift and spontaneous decay rate, respectively,
for the $i$th qubit. $\Delta_{12}$ and $\gamma_{12}$ correspond
to the exchange interaction and the mutual decay rate, respectively.
Note that $x/\lambda=1$, $1.5$, $2$, $\cdots$ correspond to nodes
of the resonant field while $x/\lambda=1.25$, $1.75$, $\cdots$
correspond to antinodes. \label{fig:shift_decay}}
\end{figure}
We start with discussion for the simplest case of two identical qubits,
who share the same frequency and bare decay rate, $\omega_{1}=\omega_{2}\equiv\omega_{0}$
and $\gamma_{12}^{0}=\gamma_{21}^{0}\equiv\gamma_{0}$, respectively.
In this case, $\Delta_{12}^{-}=\gamma_{12}^{-}=0$, $\Delta_{12}^{+}=\Delta_{12}(x_{1},x_{2})$,
$\gamma_{12}^{+}=\gamma_{12}(x_{1},x_{2})$, and $\eta_{21}=1$. Fig.
\ref{fig:shift_decay} shows the the self and cooperative decay rates
and energy shifts according to Eqs. (\ref{eq:gnm}) and (\ref{eq:dnm}).
Here, we set $x_{1}=0$, i.e., the $1$st qubit is placed at the antinode
mirror, and vary the position $x_{2}$ of the $2$nd one. Since $\gamma_{12}$
and $\Delta_{12}$ are periodic functions of $x_{2}$, we will not
lose generality if only discuss $1\le x_{2}/\lambda\le2$ with $\lambda=2\pi v/\omega_{0}$.
For decay rates shown in Fig. \ref{fig:shift_decay} (a), it can be
observed that the spontaneous ones are proportional to the strength
of the local field: $\gamma_{11}$ remains at the maximal value due
to the antinode at $x_{1}=0$, and $\gamma_{22}$ oscillates harmonically
following the intensity of the standing wave of resonant (of period
$\lambda/2$) as $x_{2}$ increases. The mutual decay rates $\gamma_{12}$
also oscillates in space but of period $\lambda$. Note that $\gamma_{12}$
changes sign every half a wavelength, suggesting the interchange of
the super- and subradiant nature between the symmetric and anti-symmetric
states. Further, in Fig. \ref{fig:shift_decay} (b) we find that the
energy shift $\Delta_{22}$ for the $2$nd qubit also displays oscillatory
features, following the same period of its KK counterpart $\gamma_{22}$.
But the amount of shift vanishes at the locations of both nodes and
antinodes ($\Delta_{11}=0$ due to $x_{1}$ fixed at the antinode).
On the other hand, the mutual coupling $\Delta_{12}$ deals with the
exchange interaction between two qubits. It is noticeable that although
$\Delta_{12}$, as the KK counterpart of $\gamma_{12}$, vanishes
at antinodes, it reaches the maximum magnitude at nodes. The nonzero
coupling mixes the symmetric and anti-symmetric states, and introduces
a finite CLS, which, we will show in the following, is visible in
the reflection spectrum.

\begin{figure}
\begin{centering}
\includegraphics[width=8.5cm]{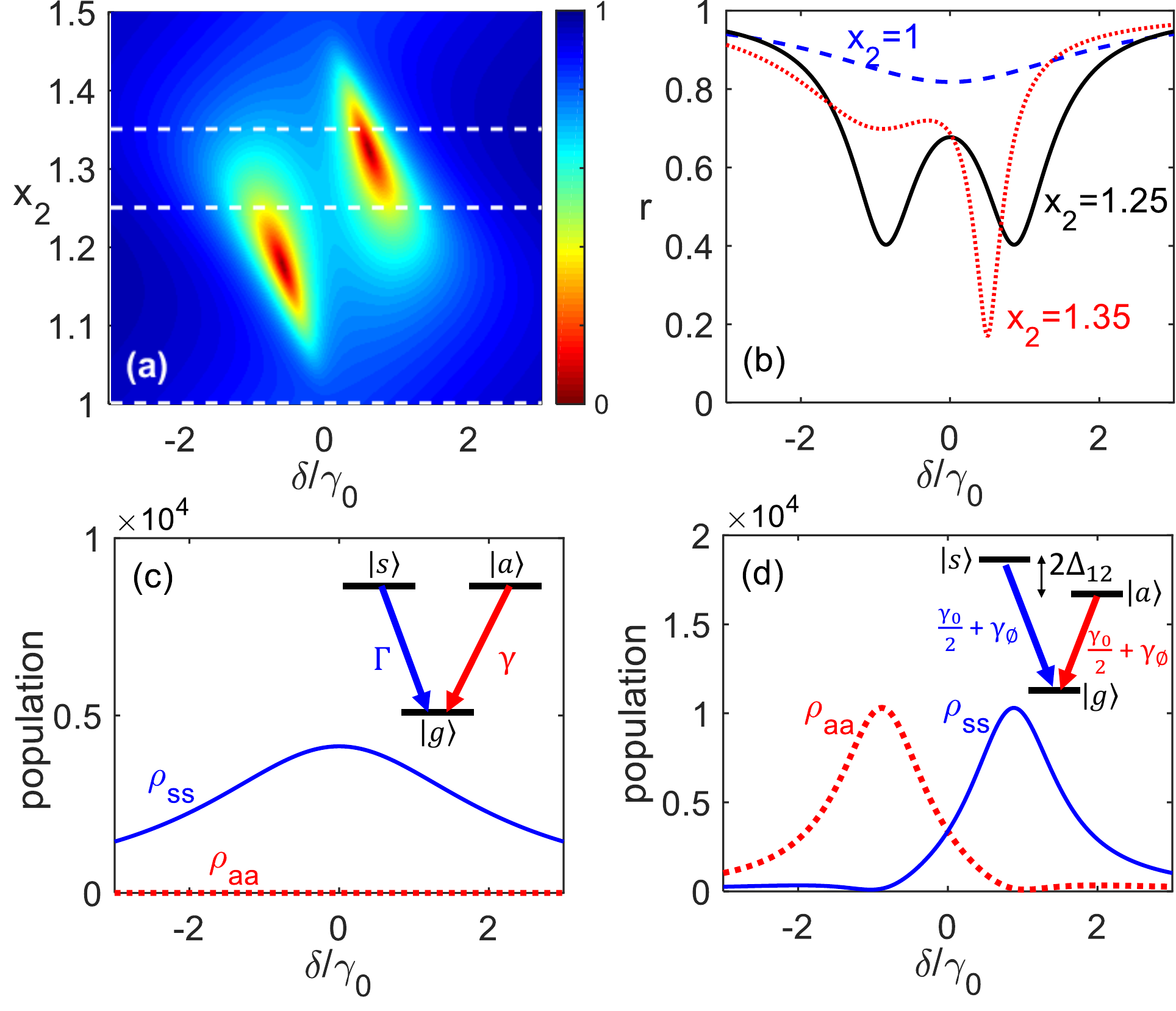}
\par\end{centering}
\caption{(a) Reflection spectrum for various $x_{2}$ in units of $\lambda$
with $x_{1}=0$. (b) The profiles corresponding to three white dashed
line cuts in (a). For $x_{2}/\lambda=1$ (antinode), the spectral
profile presents a single wide dip, signaling the superradiant nature.
For $x_{2}/\lambda=1.25$ (node), the symmetric and antisymmetric
states are split due to the CLS so that two dips merge corresponding
to two resonant conditions. For $x_{2}$ away from the antinode, two
dips move to the side of red detuning with the left one rising and
finally fading out, and the right one moving toward the middle, and
finally becoming superradiant as $x_{2}$ reaches the next antinode.
(c) Population as a function of detuning in the symmetric ($\rho_{ss}$)
and antisymmetric ($\rho_{aa}$) states for $x_{2}/\lambda=1$. (d)
Similar to (c) but for $x_{2}/\lambda=1.25$. Note that for $1.5\le x_{2}/\lambda\le2$,
these curves are similar but with the roles of the symmetric and antisymmetric
states are switched. (Other paramers: $\gamma_{\phi}=0.2\gamma_{0}$
and $\Omega_{p}=0.01\gamma_{0}$.) \label{fig:reflection}}
\end{figure}
The reflection amplitude coefficient Eq. (\ref{eq:final form reflection})
can be directly computed through solving the steady state from the
master Eq. (\ref{eq:master eq}) numerically. Fig. \ref{fig:reflection}(a)
maps out the reflection spectrum from $x_{2}/\lambda=1$ (antinode)
to $x_{2}/\lambda=1.5$ (next antinode). To understand the spectrum,
it is instructional to perform analysis under certain approximation
and circumstances. We here only briefly summarize the method we use
and the results for two qubits for (i) $x_{2}/\lambda=1$ (antinode),
and (ii) $x_{2}/\lambda=1.25$ (node) with $x_{1}=0$ in both cases.
Note that Eq. (\ref{eq:final form reflection}) can be reduced to
the standard Lindblad form \citep{Chang2012,Solano2017,LeKien2005,Guerin2017,Blais2004},
and recast into the Dicke representation. For $x_{2}/\lambda=1$,
we go to the non-Hermitian effective Hamiltonian:
\begin{equation}
\begin{aligned}H_{eff}= & -\hbar\delta(D_{s}^{+}D_{s}^{-}+D_{a}^{+}D_{a}^{-})\\
 & -\hbar\sqrt{2}\Omega_{p}(D_{s}^{+}+D_{s}^{-})\\
 & -i\hbar\Gamma D_{s}^{+}D_{s}^{-}-i\hbar\left(\frac{\gamma_{1}^{\phi}+\gamma_{2}^{\phi}}{2}\right)D_{a}^{+}D_{a}^{-}\\
 & -i\hbar\left(\frac{\gamma_{1}^{\phi}-\gamma_{2}^{\phi}}{2}\right)\left(D_{s}^{+}D_{a}^{-}-D_{a}^{+}D_{s}^{-}\right)
\end{aligned}
,\label{eq:antinode heff}
\end{equation}
where the collective symmetric and anti-symmetric operators are defined
by $D_{s}^{\pm}=\frac{1}{\sqrt{2}}(\sigma_{1}^{\pm}+\sigma_{2}^{\pm})$
and $D_{a}^{\pm}=\frac{1}{\sqrt{2}}(\sigma_{1}^{\pm}-\sigma_{2}^{\pm})$,
respectively, with the superradiant decay rate $\Gamma=2\gamma_{0}+\frac{\gamma_{1}^{\phi}+\gamma_{2}^{\phi}}{2}$.
Here, assuming the dephasing rate $\gamma_{1}^{\phi}=\gamma_{2}^{\phi}=\gamma_{\phi}$
shared by the qubits, we write the two-qubit state $|\psi\rangle=c_{g}|g\rangle+c_{s}|s\rangle+c_{a}|a\rangle+c_{e}|e\rangle$,
where $|g\rangle\equiv|gg\rangle$, $|e\rangle=|ee\rangle$, $|s\rangle=(|ge\rangle+|eg\rangle)/\sqrt{2}$,
and $|a\rangle=(|ge\rangle-|eg\rangle)/\sqrt{2}$. In the weak-field
approximation, $\Omega_{p}\ll\gamma_{0}$, $c_{g}\approx1$, we thus
omit the higher order terms $\mathcal{O}(\Omega_{p}^{2})$, and then
have $D_{s}^{+}\approx|s\rangle\langle g|$ and $D_{a}^{+}\approx|a\rangle\langle g|$.
The steady-state solution to the Schrodinger's equation $i\hbar\frac{d}{dt}|\psi\rangle=H_{eff}|\psi\rangle$
is 
\begin{align}
c_{s} & =\frac{\sqrt{2}i\Omega_{p}(\gamma_{\phi}-i\delta)}{2\gamma_{0}\gamma_{\phi}+\gamma_{\phi}^{2}-\delta^{2}-2i\delta(\gamma_{0}+\gamma_{\phi})}\\
c_{a} & =0.
\end{align}
Plugging $c_{s}$ and $c_{a}$ into Eq. (\ref{eq:final form reflection})
using $\left\langle \sigma_{1,2}^{-}\right\rangle =\frac{1}{\sqrt{2}}(c_{s}\pm c_{a})$,
we can obtain

\begin{equation}
r=\left|1-\frac{4\gamma_{0}(\gamma_{\phi}-i\delta)}{2\gamma_{0}\gamma_{\phi}+\gamma_{\phi}^{2}-\delta^{2}-2i\delta(\gamma_{0}+\gamma_{\phi})}\right|,\label{eq: antinode exact reflection}
\end{equation}
which, when $\gamma_{\phi}$ is negligible, approaches $\left|1-\frac{4\gamma_{0}\gamma_{\phi}}{\delta^{2}+4\gamma_{0}^{2}}\right|^{1/2}$
forming a central dip of width $2\gamma_{0}$. This corresponds to
the Dicke superradiant condition that the linewidth is broadened by
a factor of $2$ for two qubits. Also, we can see that only the symmetric
state $|s\rangle$ is occupied with population $\rho_{ss}=|c_{s}|^{2}$
because the anti-symmetric state is decoupled. Note that when $x_{2}/\lambda=1.5$,
the roles of the symmetric and antisymmetric states are switched because
$\gamma_{12}=-\gamma_{0}$ and $\sigma_{1}^{\pm}+(-1)^{2x_{2}/\lambda}\sigma_{2}^{\pm}\rightarrow\sqrt{2}D_{a}^{+}$,
and also $\Gamma\rightarrow\gamma_{\phi}$ and $\gamma\rightarrow2\gamma_{0}+\gamma_{\phi}$
in Eq. (\ref{eq:antinode heff}). 

For $x_{2}=1.25\lambda$, the Hamiltonian then becomes
\begin{equation}
\begin{aligned}H_{eff}= & \hbar(-\delta+\Delta_{12})D_{s}^{+}D_{s}^{-}\\
 & +\hbar(-\delta-\Delta_{12})D_{a}^{+}D_{a}^{-}\\
 & -\frac{\hbar\Omega_{p}}{\sqrt{2}}(D_{s}^{+}+D_{s}^{-}+D_{a}^{+}+D_{a}^{-})\\
 & -i\hbar\gamma^{+}(D_{s}^{+}D_{s}^{-}+D_{a}^{+}D_{a}^{-})\\
 & -i\hbar\gamma^{-}(D_{s}^{+}D_{a}^{-}-D_{a}^{+}D_{s}^{-})
\end{aligned}
\label{eq:node heff}
\end{equation}
with
\begin{equation}
\gamma^{\pm}=\frac{\gamma_{0}+\gamma_{\phi_{1}}\pm\gamma_{\phi_{2}}}{2}.\label{eq: node case ga}
\end{equation}
Similar analysis leads to the solution
\begin{align}
c_{s} & =\frac{\frac{i\Omega_{p}}{\sqrt{2}}\left[\gamma_{2}^{\phi}-i(\delta+\Delta_{12})\right]}{(\gamma_{0}+\gamma_{1}^{\phi})\gamma_{2}^{\phi}-(\delta^{2}-\Delta_{12}^{2})-2i\delta\gamma^{+}}\label{eq:exact cs}\\
c_{a} & =\frac{\frac{i\Omega_{p}}{\sqrt{2}}\left[\gamma_{2}^{\phi}-i(\delta-\Delta_{12})\right]}{(\gamma_{0}+\gamma_{1}^{\phi})\gamma_{2}^{\phi}-(\delta^{2}-\Delta_{12}^{2})-2i\delta\gamma^{+}}\label{eq:exact ca}
\end{align}
and hence
\begin{equation}
r=\left|1-\frac{2\gamma_{0}(\gamma_{2}^{\phi}-i\delta)}{(\gamma_{0}+\gamma_{1}^{\phi})\gamma_{2}^{\phi}-(\delta^{2}-\Delta_{12}^{2})-2i\delta\gamma^{+}}\right|.\label{eq: node exact reflection}
\end{equation}
For small $\gamma_{2}^{\phi}$, two dips correspond to $\delta\rightarrow\delta_{\pm}$
with
\begin{align}
\delta_{\pm} & \approx\pm\Delta_{12}\left[1-\frac{\gamma_{0}^{2}-\gamma_{1}^{\phi2}}{4\Delta_{12}^{2}}\frac{\gamma_{2}^{\phi}}{\gamma_{1}^{\phi}}\right]\rightarrow\pm\Delta_{12}\label{eq:delta_pm}
\end{align}
as $\gamma_{2}^{\phi}\rightarrow0$. This is consistent with our previous
argument that $\Delta_{12}$ contributes to a coupling between $|s\rangle$
and $|a\rangle$ and splits the two states. Therefore, the CLS directly
results in the spectral splitting $\delta_{split}\equiv2|\delta_{\pm}|\approx2\Delta_{12}$
directly evident in the the reflection profile. Finally, note that
at $\delta=-\Delta_{12}$, $\rho_{ss}\approx\frac{|\Omega_{p}|^{2}\gamma_{2}^{\phi2}}{2\Delta_{12}^{2}\gamma_{0}^{2}}\rightarrow0$
as $\gamma_{2}^{\phi}\rightarrow0$, implying that all the excitation
is populated on state $|a\rangle$. Conversely, at $\delta=+\Delta_{12},$only
state $|s\rangle$ is populated. In the case of $x_{2}/\lambda=1.75$,
the roles of the symmetric and antisymmetric states are switched due
to the same argument in the case of $x_{2}/\lambda=1.5$ discussed
previously.

\subsection{Dephasing\label{subsec:dephasing}}

\begin{figure}
\begin{centering}
\includegraphics[width=8.5cm]{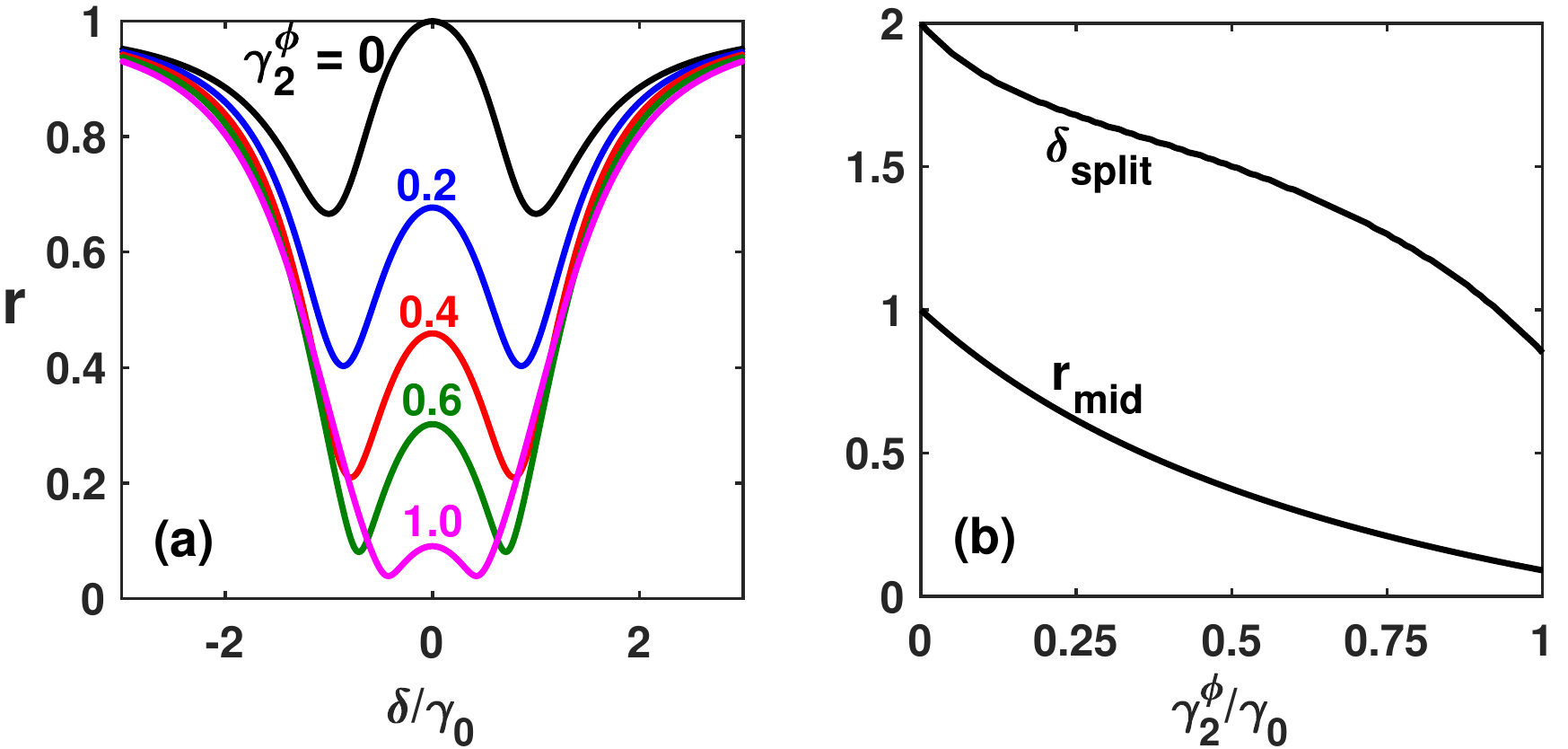}
\par\end{centering}
\caption{(a) Reflection spectrum for various dephasing rates of the $2$nd
qubit at $x_{2}/\lambda=1.25$. Here we set $\gamma_{1}^{\phi}=0.2\gamma_{0}$
and $\Omega_{p}=0.01\gamma_{0}$. (b) Spectral splitting $\delta_{split}$
in units of $\gamma_{0}$ and the height of the central maximum $r_{mid}$
as monotonically descending functions of the 2nd qubit's dephasing
rate $\gamma_{2}^{\phi}$. \label{fig:dephasing}}
\end{figure}
We now examine the effect of dephasing on the splitting feature. Intuitively
speaking, dephasing usually introduces broadening that degrades the
quantum effects from being observed. In our case, however, when we
take $\gamma_{1}^{\phi}=\gamma_{2}^{\phi}=0$, Eq. (\ref{eq: node exact reflection})
gives $r=1$ constant reflection profile for any finite detuning $\delta$.
Therefore the splitting information will be hidden. In fact, we need
$\gamma_{1}^{\phi}>0$ in order to view splitting as a trace of CLS
from the reflection spectrum. We have shown in Eq. (\ref{eq:delta_pm})
that $\delta_{\pm}\rightarrow\pm\Delta_{12}$ as $\gamma_{2}^{\phi}\rightarrow0$
for any $\gamma_{1}^{\phi}>0$. When $\gamma_{2}^{\phi}>0$, we find
that the mismatch between $\delta_{split}$ and $2\Delta_{12}$ has
a leading-order term proportional to $\gamma_{2}^{\phi}/\gamma_{1}^{\phi}$,
which suggests that $\delta_{\pm}\rightarrow\pm\Delta_{12}$ as long
as $\gamma_{2}^{\phi}/\gamma_{1}^{\phi}$ is small.

Figure \ref{fig:dephasing} shows our numerical calculation when $\gamma_{1}^{\phi}=0.2\gamma_{0}$
is fixed, corresponding to a typical experimental realization. When
$\gamma_{2}^{\phi}$ increases from zero, we find $\delta_{split}$
decreases monotonically from $2\Delta_{12}$. Another interesting
feature regarding visibility of CLS is the central maximum $r_{mid}\equiv r(\delta=0)$,
which is also lowered with increasing $\gamma_{2}^{\phi}$ according
to 
\begin{equation}
r_{mid}=1-\frac{2\gamma_{11}\gamma_{2}^{\phi}}{(\gamma_{11}+\gamma_{1}^{\phi})\gamma_{2}^{\phi}+\Delta_{12}^{2}}.\label{eq:r_mid}
\end{equation}
In real experiments \citep{Wen2019}, this maximum is always smaller
than unity, reflecting the presence of dephasing mechanisms on the
$2$nd qubit. We find that $r_{mid}$ is dominantly determined by
$\gamma_{2}^{\phi}$ and insensitive to $\gamma_{1}^{\phi}$ according
to Eq. (\ref{eq:r_mid}). Thus $r_{mid}$ provides a very good indication
to be used to extract $\gamma_{2}^{\phi}$ by neglecting $\gamma_{1}^{\phi}$.
The ratio of $\gamma_{2}^{\phi}$ thus obtained to the actual value
is $\Delta_{12}^{2}/(\Delta_{12}^{2}+\gamma_{1}^{\phi}\gamma_{2}^{\phi})$.
Therefore, for $\gamma_{1}^{\phi}$, $\gamma_{2}^{\phi}\sim0.5\gamma_{0}$,
the estimated value of $\gamma_{2}^{\phi}$ is 20\% less than the
actual one; for $\gamma_{1}^{\phi}$, $\gamma_{2}^{\phi}\sim0.2\gamma_{0}$
as in a typical experiment, it becomes only 4\% less.

\subsection{Power broadening }

\begin{figure}
\noindent \begin{centering}
\includegraphics[width=6.5cm]{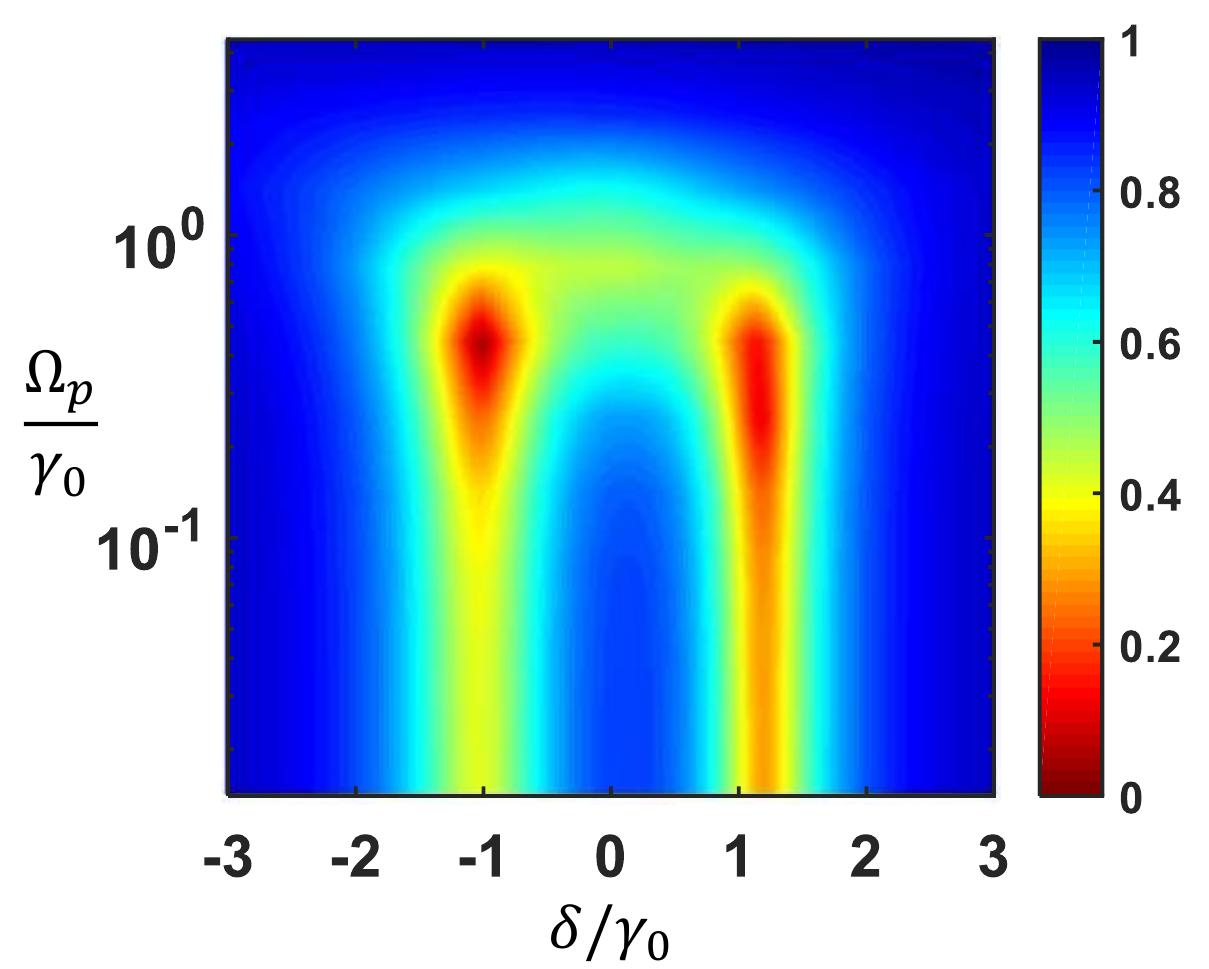}
\par\end{centering}
\caption{Power broadening of the reflection spectrum of one qubit located at
$x_{1}=0$ and one at $x_{2}=1.755\lambda$, slightly off the node
of the standing wave. See text for other parameters. \label{fig:power broadening}}
\end{figure}
In preceding discussion, we have focused on the weak field limit on
two identical qubits with one located at the anti-node and the other
at the node perfectly. In this section, we discuss a more realistic
case with parameters given in the experiment \citep{Wen2019}. For
strong probing, the effective-Hamiltonian approach breaks down at
some point due to significant population in the upper level. To account
for the associated effect, we generalize our model to multi-level
atoms (see Appendix B), where the anharmonicity of transmon qubits
is explicitly considered, and numerically solve the master equation
Eq. (\ref{eq:app_many master eq}). A typical power dependent reflection
spectrum is shown in Fig. \ref{fig:power broadening}. Here, we have
used $\omega_{1}=2\pi\times4.755$ GHz and $\omega_{2}=2\pi\times4.759$
GHz for two atoms at $x_{1}=0$ and $x_{2}=1.755\lambda$, respectively,
where $\lambda=2\pi v/\omega_{2}$ with wavespeed $v=0.8948\times10^{8}$
m/s . Further, $\gamma_{11}\approx\gamma_{12}^{0}\approx\gamma_{21}^{0}\approx\gamma_{0}=2\pi\times17.2$
MHz, $\gamma_{22}\approx2\pi\times0.02$ MHz, the anharmonicity is
$2\pi\times406$ MHz and $2\pi\times324$ MHz for the $1$st and $2$nd
atoms, respectively. The dephasing rates are $0.17\gamma_{0}$ ($0.28\gamma_{0}$)
and $0.13\gamma_{0}$ ($0.25\gamma_{0}$) for the lower (upper) level
transition of the $1$st and $2$nd atoms, respectively.

For weak probing $\Omega_{p}\lesssim0.1\gamma_{0}$, the spectrum
profiles remain independent of the probe power, reflecting the fact
that the CLS originates from vacuum modes rather than the real photon
field. As $\Omega_{p}$ increases, we see clear power broadening of
the two dips due to significant population in the second level. Note
that the spectral profile being slightly shifted to the right is because
of frequency mismatch of the two qubits. And the obvious asymmetry
of the profile is owing to the fact that the $2$nd qubit is not perfectly
placed at the node and contributes to the scattered signal. Though
our model explicitly considers the third level, our numerical results
suggest that its role is almost negligible as long as the anharmonicity
is greater than $5\gamma_{0}\approx2\pi\times87$ MHz. For $\Omega_{p}\gtrsim2\gamma_{0}$,
the system gets saturated and attains unity reflection.

\section{Multi-atom cases\label{sec:manyatom}}

\begin{figure}
\begin{centering}
\includegraphics[width=7.5cm]{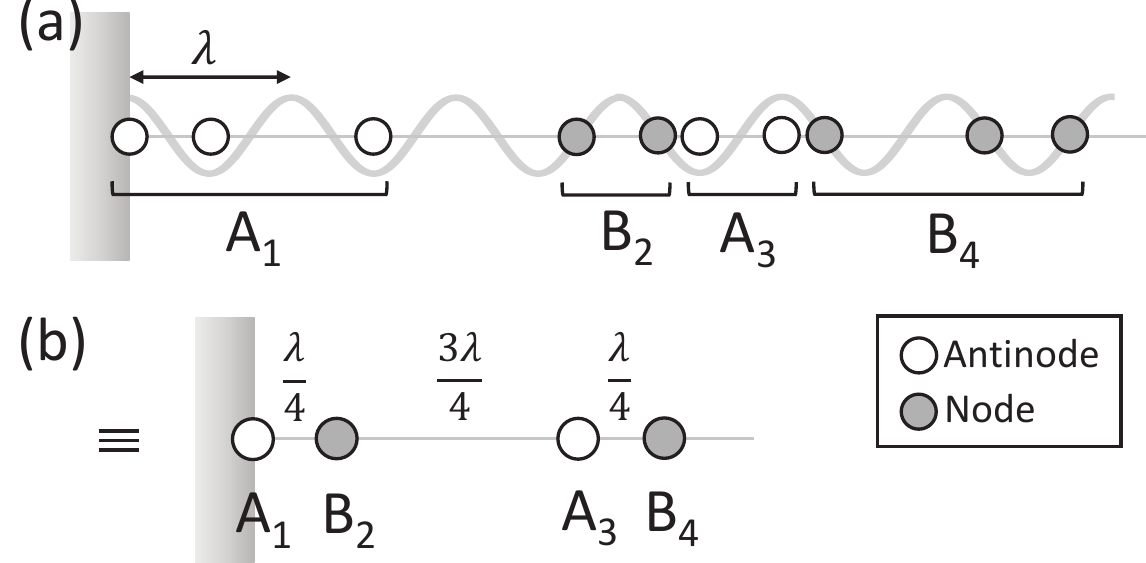}
\par\end{centering}
\caption{(a) Array of qubits located at either nodes and antinodes. (b) Equivalent
reduced scheme of ``giant atoms'' arranged at antinodes and nodes
alternatively. \label{fig:atomarray}}
\end{figure}
\begin{figure}
\begin{centering}
\includegraphics[width=8.5cm]{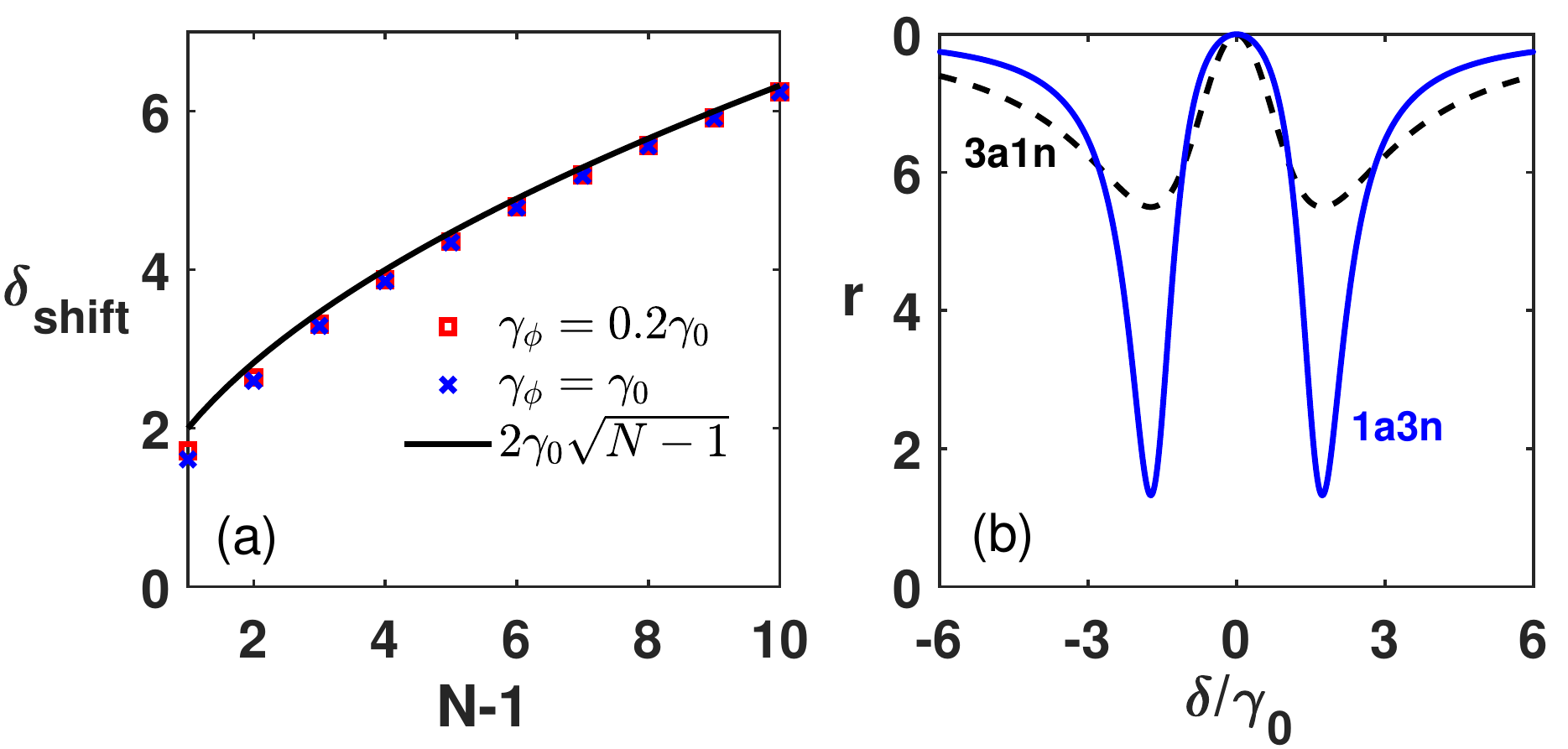}
\par\end{centering}
\caption{(a) CLS splitting $\delta_{split}\approx2\gamma_{0}\sqrt{N-1}$ for
an qubit array of one at the mirror ($x_{1}=0$) and $N-1$ ones at
nodes. Small deviations can be observed with finite dephasing rate
$\gamma_{\phi}$ for all the qubits. (b) Spectral profiles for three
antinode qubits plus one node qubit (3a1n) and one antinode qubit
plus three node ones (1a3n). \label{fig:scaling}}
\end{figure}
We now consider multi-atom cases with $N\ge3$. Although the system's
dynamics can be fully determined by evolving the master equation (\ref{eq:master eq})
for arbitrary arrangements of qubits, we here focus on configurations
with identical qubits either at antinodes or nodes as shown in Fig.
\ref{fig:atomarray} (a). For analysis, we first take those qubits
at antinodes/nodes in a row as a group. By doing so, the system now
consists of antinode groups ($A_{j}$) and node ones ($B_{j}$) placed
in alternative order, i.e., $A_{1}B_{2}A_{3}B_{4}\cdots$. For each
antinode group, we define the collective operator as $S_{j}^{\pm}\equiv\sum_{i\in A_{j}}(-1)^{2x_{i}/\lambda}\sigma_{i}^{\pm}$,
and for each node one $S_{j}^{\pm}=\sum_{i\in B_{j}}(-1)^{2x_{i}/\lambda-0.5}\sigma_{i}^{\pm}$.
Under the weak field approximation with only single excitation allowed,
$S_{j}^{+}\approx|s\rangle_{A_{j}}\langle g|^{\otimes n_{j}}$ with
$|s\rangle$ a superradiant state. For instance, for the case $(x_{1},x_{2},x_{3})=(0,1,1.5)\lambda$,
$S^{+}=\sigma_{1}^{+}+\sigma_{2}^{+}-\sigma_{3}^{+}$, the superradiant
state is given by $|s\rangle=\frac{1}{\sqrt{3}}(|e,g,g\rangle+|g,e,g\rangle-|e,e,g\rangle)$.
Thus the group $A_{j}$ can be seen as a ``giant'' atom effectively
with only two levels $|g\rangle^{\otimes n_{j}}$ and $|s\rangle$.
The reason why $|s\rangle$ is superradiant can be seen by noticing
the dipole-dipole interaction (in the non-Hermitian representation)
characterized by the decay terms $\sum_{ii^{\prime}}\gamma_{ii^{\prime}}\sigma_{i}^{+}\sigma_{i^{\prime}}^{-}\rightarrow\sum_{A_{j}}n_{j}\gamma_{0}S_{j}^{+}S_{j}^{-}+\sum_{A_{j},A_{j^{\prime}}}\sqrt{n_{j}n_{j^{\prime}}}\gamma_{0}S_{j}^{+}S_{j^{\prime}}^{-}$,
where the first terms correspond to superradiant decay of $A_{j}$,
and the second terms correspond to mutual decay channels between different
giant atoms $A_{j}$ and $A_{j^{\prime}}$. There are no mutual decay
terms between $A_{j}$ and $B_{j^{\prime}}$, and between $B_{j}$
and $B_{j^{\prime}}$. Similarly, the dipole-dipole interaction characterized
by exchange $\sum_{ij}\Delta_{ij}\sigma_{i}^{+}\sigma_{j}^{-}\rightarrow\sum_{A_{j},B_{j^{\prime}},j<j^{\prime}}\sqrt{n_{j}n_{j^{\prime}}}\gamma_{0}S_{j}^{+}S_{j^{\prime}}^{-}$.
We can also treat $B_{j^{\prime}}$ as a giant two-level atom in the
same manner, we find the exchange coupling only exists between $A_{j}$
and $B_{j^{\prime}}$ when $j<j^{\prime}$. For $j>j^{\prime}$, from
Eq. (\ref{eq:dnm}) where $\Delta_{ii^{\prime}}\sim\sin k_{0}(x_{i}+x_{i^{\prime}})+\sin k_{0}|x_{i}-x_{i^{\prime}}|$,
the two sine terms differ by a phase of $\pi$ and hence cancel out.
This leads us to an effective reduced scheme represented by Fig. \ref{fig:atomarray}
(b), which will yield the same spectral behavior as the original one.

In the case of an array consisting of two groups $A_{1}$ and $B_{2}$,
with $n_{1}$ antinode and $n_{2}$ node qubits, respectively, it
can be directly replaced by one antinode atom of linewidths $\sqrt{n_{1}}\gamma_{0}$
coupled to a node atom with exchange coupling $\sqrt{n_{1}n_{2}}\gamma_{0}$.
Thus, the CLS splitting is $\delta_{split}=2\sqrt{n_{1}n_{2}}\gamma_{0}$
without dephasing. Fig. \ref{fig:scaling} (a) presents the scaling
law of $\delta_{split}$, which indeed agrees with our prediction.
Small deviation is visible but negligible when dephasing is included,
and gradually vanishes as $N$ gets large. In Fig. \ref{fig:scaling}
(b) we compare the reflection spectral profile of two situations:
$(x_{1},x_{2},x_{3},x_{4})=(0,1,2,2.25)\lambda$ and $(0,0.25,0.75,1.25)\lambda$.
In both cases the splitting is the same $2\sqrt{3}\gamma_{0}$. But
the former has a broadened linewidth $\sqrt{3}\gamma_{0}$ due to
superradiant enhancement in $A_{1}$.

\section{Conclusion\label{sec:conclusion}}

In summary, we have studied the dipole-dipole interaction between
artificial atoms mediated by 1D vacuum modes in a waveguide. Setting
one end of the waveguide to be a mirror, we are allowed to probe the
collective Lamb shift by studying the reflection spectrum. When an
qubit is placed at the node, we isolate it from coupling to other
qubits through real photon field. Instead, the exchange interaction
via virtual photons remains at work, causing the collective Lamb splitting
between symmetric and antisymmetric levels that can now be clearly
visible by means of a very simple reflection measurement. Our calculation
highly agrees with the experimental results in \citep{Wen2019}. We
have derived the master equation to describe general cases and given
analytical expressions for certain circumstances. We have also investigated
the effects of dephasing, and the scaling law when more qubits are
added. For special cases with many qubits placed only at antinodes
and nodes, we have developed a reduced scheme under the weak field
approximation, and explained the scaling behavior. For future outlook,
we find close connection of our findings to a recent work \citep{Kockum2018},
where atoms are considered large compared to the transition wavelength,
and thought to have multiple chances of interaction before the field
leaves. We expect similar analysis for some interesting interference
effects, and our results can be very useful for quantum optical study
and quantum simulation.
\begin{acknowledgments}
K.-T. L. acknowledges Zih-Sin Chan and Yun-Chih Liao for helpful discussion.
I.-C. H. acknowledges financial support from the MOST of Taiwan under
Grant No. 107-2112-M-007-008-MY3. G.-D. L. acknowledges Anton Frisk
Kockum for helpful feedback, and also the support from the MOST of
Taiwan under Grant No. 105-2112-M-002-015-MY3 and National Taiwan
University under Grant No. NTUCC-108L893206.
\end{acknowledgments}

\section*{appendix}

\subsection{Derivation of dipole-dipole interaction \label{subsec:appa_rddi}}

\setcounter{equation}{0} \renewcommand{\theequation}{A.\arabic{equation}}

In this section, we give detailed derivation of the dipole-dipole
interaction, Eqs. (\ref{eq:gnm}) and (\ref{eq:dnm}), in a 1D waveguide
and the master equation (\ref{eq:master eq}) in the main text. According
to the Hamiltonian (\ref{eq:interaction}), the equations of motion
for the field operator $a_{\omega}$ and an arbitrary atomic observable
$Q$ of the system are given respectively by 
\begin{equation}
\dot{a}_{\omega}=-i\omega a_{\omega}-\sum_{i=1}^{N}g_{i}(\omega)\cos(k_{\omega}x_{i})\sigma_{i}^{-}(t)\label{eq:app_dynamics of a}
\end{equation}
and
\begin{align}
\dot{Q}= & -\frac{i}{\hbar}\sum_{i=1}^{N}[Q,H_{s}]\nonumber \\
 & +\sum_{i=1}^{N}\int d\omega g_{i}(\omega)\cos(k_{\omega}x_{i})[Q,a_{\omega}\sigma_{i}^{+}]+H.c..\label{eq:app_dynamics of q}
\end{align}
Integrating Eq. (\ref{eq:app_dynamics of a}), we obtain
\begin{align}
a_{\omega}(t)= & a_{\omega}(0)e^{-i\omega t}\nonumber \\
 & -\int_{0}^{t}dt^{\prime}e^{i\omega(t^{\prime}-t)}\sum_{i=1}^{N}g_{i}(\omega)\cos(k_{\omega}x_{i})\sigma_{i}^{-}(t^{\prime}),
\end{align}
which, under the weak coupling and Born-Markov approximation, becomes
\begin{equation}
a_{\omega}(t)=-\sum_{i=1}^{N}\frac{g_{i}(\omega)\cos(k_{\omega}x_{i})}{\epsilon+i(\omega-\omega_{i})}\sigma_{i}^{-}(t)+\xi_{\omega}.\label{eq:app_solution of a}
\end{equation}
Here, we have used $\sigma_{i}^{-}(t^{\prime})\approx\sigma_{i}^{-}(t)e^{i\omega_{i}(t-t^{\prime})}$
and taken the upper limit of the integral $t\rightarrow\infty$. Also,
the term regarding $a_{\omega}(0)$ contributes to a noise operator
averaged out in the vacuum state. Note that the insertion of a small
positive quantity $\epsilon$ is to guarantee the convergence of the
integration. Substituting Eq. (\ref{eq:app_solution of a}) into Eq.
(\ref{eq:app_dynamics of q}), we get 
\begin{equation}
\dot{Q}=-\frac{i}{\hbar}\sum_{i=1}^{N}[Q,H_{s}]-\sum_{i,j=1}^{N}F_{ij}(\omega,x)[Q,\sigma_{i}^{+}]\sigma_{j}^{-}+H.c.\label{eq:app_qt}
\end{equation}
with 
\begin{equation}
F_{ij}(\omega,x)=\int_{0}^{\infty}\frac{g_{i}(\omega)g_{j}(\omega)\cos(k_{\omega}x_{i})\cos(k_{\omega}x_{i})}{\epsilon+i(\omega-\omega_{j})}d\omega\label{eq:app_ddi}
\end{equation}
accounting for the dipole-dipole interaction. Thus, the decay rates
and energy shift correspond to the real and imaginary part of $F_{ij}\equiv\gamma_{ij}+i\Delta_{ij}$,
respectively \citep{Lehmberg1970}. Using the technique of contour
integration, we finally obtain
\begin{align}
\gamma_{ij} & =\frac{\gamma_{ij}^{0}}{2}\left[\cos k_{j}(x_{i}+x_{j})+\cos k_{j}\left|x_{i}-x_{j}\right|\right]\label{eq:app_gij}\\
\Delta_{ij} & =\frac{\gamma_{ij}^{0}}{2}\left[\sin k_{j}(x_{i}+x_{j})+\sin k_{j}\left|x_{i}-x_{j}\right|\right]\label{eq:app_dij}
\end{align}
where $\gamma_{ij}^{0}\equiv\pi g_{i}(\omega_{j})g_{j}(\omega_{j})$.
Note that in the most general case with non-identical qubits, $\gamma_{ij}^{0}\neq\gamma_{ji}^{0}$.

\subsection{Multi-level system\label{subsec:appb_multilevel}}

This section summarizes the general form for multi-level atoms of
the master equation and dipole-dipole interaction among transmon qubits
in the setup of this work. When a collection of such atoms with states
$\left\{ |n\rangle_{i}\right\} $ ($i=1,2,\cdots,N$ is the atom index;
$n=0,1,\cdots$ is the state index) pumped by a single-mode field,
the interaction Hamiltonian can be described by 

\begin{align}
H_{int}= & i\hbar\sum_{i,n}\sqrt{n}g_{i}(\omega)\cos\frac{\omega}{c}x_{i}\left[a_{\omega}\sigma_{n,n-1}^{i}-H.c.\right],\label{eq:Hint_many}
\end{align}
where $\sigma_{n,m}^{i}\equiv|n\rangle_{i}\langle m|$ is the jumping
operator taking the $i$th atom from state $|m\rangle_{i}$ to state
$|n\rangle_{i}$. Through the standard procedure discussed in the
previous section, we derive the master equation that reads

\begin{equation}
\begin{aligned}\frac{d\rho}{dt} & =i\sum_{i,n}[\delta_{n}^{i}|n\rangle_{i}\langle n|,\rho]\\
 & +i\sum_{i,n}\sqrt{n}\Omega_{p}^{i}[\sigma_{n,n-1}^{i}+H.c.,\rho]\\
 & -i\sum_{i,j,m,n}\left\{ \frac{(\Delta_{ij,m}+\Delta_{ji,n})-i(\gamma_{ij,m}-\gamma_{ji,n})}{2}\right\} \times\\
 & \sqrt{nm}[\sigma_{n,n-1}^{i}\sigma_{m-1,m}^{j},\rho]\\
 & +\sum_{ij,m,n}\left\{ \frac{(\gamma_{ij,m}+\gamma_{ji,n})+i(\Delta_{ij,m}-\Delta_{ji,n})}{2}\right\} \times\\
 & \sqrt{nm}\mathcal{L}_{n,n-1;m-1,m}^{ij}[\rho]\\
 & +\sum_{i,n}\gamma_{i,n}^{\phi}D_{i,n}[\rho]
\end{aligned}
\label{eq:app_many master eq}
\end{equation}
with the decay rates and energy shifts
\begin{align}
\gamma_{ij,n}= & \frac{\pi g_{i}(\omega_{n}^{j})g_{j}(\omega_{n}^{j})}{2}\times\nonumber \\
 & \left[\cos\frac{\omega_{n}^{j}}{c}(x_{i}+x_{j})+\cos\frac{\omega_{n}^{j}}{c}\left|x_{i}-x_{j}\right|\right]\label{eq:app_many gij}\\
\Delta_{ij,n}= & \frac{\pi g_{i}(\omega_{n}^{j})g_{j}(\omega_{n}^{j})}{2}\times\nonumber \\
 & \left[\sin\frac{\omega_{n}^{j}}{c}(x_{i}+x_{j})+\sin\frac{\omega_{n}^{j}}{c}\left|x_{i}-x_{j}\right|\right]\label{eq:app_many dij}
\end{align}
respectively, associated with state $|n\rangle$. Also, the incoherent
jumping and dephasing superoperators are $\mathcal{L}_{mn,m^{\prime}n^{\prime}}^{ij}[\rho]_{nm}=2\sigma_{m^{\prime}n^{\prime}}^{j}\rho\sigma_{mn}^{i}-\rho\sigma_{mn}^{i}\sigma_{m^{\prime}n^{\prime}}^{j}-\sigma_{mn}^{i}\sigma_{m^{\prime}n^{\prime}}^{j}\rho$
and $D_{i,n}[\rho]=2\sigma_{nn}^{i}\rho\sigma_{nn}^{i}-\rho\sigma_{nn}^{i}-\sigma_{nn}^{i}\rho$,
respectively. By solving the master equation, one can obtain the reflection
coefficient through

\begin{equation}
r=\left|1+i2\frac{Q^{N}}{\Omega_{01}^{N}}\sum_{i,n}\frac{\gamma_{ii,n}}{Q^{i}}\sqrt{n}\cos k_{p}x_{i}\langle\sigma_{n-1,n}^{i}\rangle\right|,\label{eq:app_many reflection}
\end{equation}
where $Q^{n}=\sqrt{2}e\beta_{n}(E_{J}^{(n)}/8E_{c}^{(n)})^{1/4}$.

\bibliographystyle{apsrev4-1}
\bibliography{cls_cqed_20190509}

\end{document}